\documentclass[%
 aip,
 amsmath,amssymb,
 reprint,%
]{revtex4-1}

\usepackage{graphicx}
\usepackage{dcolumn}
\usepackage{bm}

\usepackage[utf8]{inputenc}
\usepackage[T1]{fontenc}
\usepackage{mathptmx}
\usepackage{etoolbox}
\usepackage{upgreek}

\usepackage{xcolor}

\definecolor{redline}{rgb}{0.8,0,0.0.5}

\makeatletter
\def\@email#1#2{%
 \endgroup
 \patchcmd{\titleblock@produce}
  {\frontmatter@RRAPformat}
  {\frontmatter@RRAPformat{\produce@RRAP{*#1\href{mailto:#2}{#2}}}\frontmatter@RRAPformat}
  {}{}
}%
\makeatother
\begin{document}

\preprint{AIP/123-QED}

\title[Nonlinear Excitation of Ideal Anapoles]{Nonlinear Excitation of Ideal Anapoles}
\author{R. Kolkowski}

\author{M. Kaivola}
 
\author{A. Shevchenko}
\email{radoslaw.kolkowski@aalto.fi, andriy.shevchenko@aalto.fi.}

\affiliation{%
Department of Applied Physics, Aalto University, P.O.Box 13500, Aalto FI-00076, Finland
}%

\date{\today}

\begin{abstract}
Optical anapoles have attracted significant interest as a promising platform for enhancing light-matter interactions at the nanoscale. An ideal anapole is a localized excitation of electromagnetic fields that does not produce any far-field radiation. In practice, anapoles are not ideal due to the parasitic scattered fields associated with higher-order multipoles. Here, we use the recently established \emph{exact current multipole expansion} and the \emph{exact anapole condition} to design an ideal anapole in which radiation due to the multipole orders up to the electric and magnetic octupoles is either eliminated or strongly suppressed, reducing the total radiated power by orders of magnitude. We also propose and study realistic photonic structures in which such anapoles can be excited by a uniform linearly polarized plane wave through second-harmonic generation that involves the off-diagonal elements of the second-order susceptibility tensor. In this case, the anapole condition is reached off-resonantly by adjusting the polarization angle of the incident field, which allows for the anapoles to be excited over a wide spectral range in scatterers of different sizes. Using this approach, we achieve scattering suppression on the order of $10^{-4}$ in photonic structures that are half a wavelength in size. The method offers the possibility to realize localized electromagnetic excitations that are truly isolated from the surroundings.
\end{abstract}

\maketitle

\section{\label{sec:intro}Introduction}

Efficient confinement of light at the nanoscale is one of the central goals in nanophotonics, enabling enhancement of light-matter interactions and realization of compact photonic devices such as lasers, sensors, and nonlinear optical components~\cite{vahala03,koenderink15,wu21,koshelev21}. In photonic micro- and nanoresonators, the achievable local intensity enhancement and quality ($Q$) factors are limited by the radiative loss. Therefore, suppressing this type of loss, e.g., via destructive interference in the far field, is an important task in nanophotonics, with a notable example of photonic bound states in the continuum~\cite{hsu16,koshelev19,koshelev19_,kang23}. In the case of anapoles, such destructive interference occurs between two different types of excitations, e.g., a dipole and a toroidal dipole, which produce identical but out-of-phase radiation patterns~\cite{miroshnichenko15,savinov19,baryshnikova19,yang19}. Unfortunately, many practical realizations of anapoles (using, for example, dielectric nanodiscs) are not perfect, exhibiting non-negligible radiation by higher-order multipoles, such as magnetic quadrupoles~\cite{wei16,ospanova20,canos21}. Suppressing this parasitic scattering, e.g., by structuring the incident field~\cite{wei16}, has been proposed previously, but practical realization of a perfectly nonradiating anapole remains elusive. 

\begin{figure}[t]
\includegraphics[width=85mm]{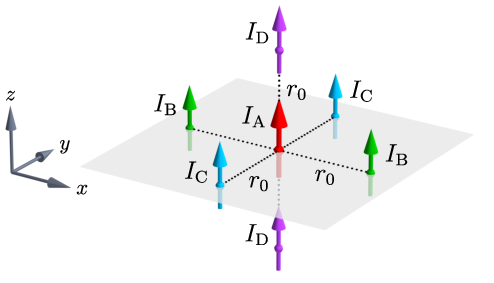}
\caption{\label{fig:1} Illustration of the scattering current configurations considered in this work. In general, each configuration consists of seven current elements, one with amplitude $I_{\text{A}}$ placed at the origin of the coordinate system and six with amplitudes $I_{\text{B}}$, $I_{\text{C}}$, and $I_{\text{D}}$ positioned on the Cartesian axes at a distance $r_0$ from the origin. The vectors of the scattering currents are pointing either along $z$ or along $-z$ depending on the signs of $I_{\text{A}}$, $I_{\text{B}}$, $I_{\text{C}}$, and $I_{\text{D}}$.}
\end{figure}

In this work, we use the \emph{exact scattering-current multipole expansion}~\cite{kolkowski26} to design ideal anapoles composed of several parallel current elements, as illustrated in Fig.~\ref{fig:1}. In the figure, the amplitudes of the currents at the coordinates of these elements are defined by $I_i$, $i\in\{\textrm{A, B, C, D}\}$. The current density in the elements can be written in terms of the three-dimensional Dirac delta function as $J_i=I_i L\delta(\textbf{r}\pm\textbf{r}_i)$, where $\textbf{r}_i$ is the coordinate of the element and $L$ is its infinitesimally small length. Using the scattering-current multipole expansion, we derive the \emph{exact anapole condition} for this configuration, which can be expressed as (see Section~\ref{sec:design})
\begin{multline}\label{eq:eac0}
\qquad I_{\text{A}}=\left(I_{\text{B}}+I_{\text{C}}\right)\left[j_2(kr_0)-2j_0(kr_0)\right]\\-2I_{\text{D}}\left[j_2(kr_0)+j_0(kr_0)\right],\qquad
\end{multline}
where $j_n(kr_0)$ are the spherical Bessel functions of the first kind of order $n$, $k$ is the wavenumber in the surrounding medium, and $r_0$ is the distance of the peripheral current elements from the central element (see Fig.~\ref{fig:1}). An ideal anapole is obtained by setting 
\begin{equation}\label{eq:setting}
I_{\text{B}}=I_{\text{C}}=I_{\text{D}}=-\frac{I_{\text{A}}}{6j_0(kr_0)},\end{equation} 
which leads to complete elimination of scattering due to electric and magnetic dipoles and quadrupoles, as well as all higher-order electric multipoles of even order and magnetic multipoles of odd order. At the same time, the electric-octupole scattering is strongly suppressed. The lowest-order unsuppressed contribution comes from the magnetic hexadecapoles, which become a dominant source of radiation. As a result, for wavelength-sized scatterers, the radiated power is reduced to $\sim$2\% of the power radiated by a point dipole with an equivalent current. For subwavelength-sized scatterers, the radiated power drops by several orders of magnitude, turning them into perfectly dark anapole scatterers. This current configuration has previously been discovered in the current multipole expansion as one of the anapoles composed of current multipole moments of different orders~\cite{grahn12,shevchenko20}.

Inspired by these findings, we design and numerically demonstrate realistic photonic nanostructures that support ideal anapoles. Their excitation is possible under uniform plane-wave illumination (in the ideal case requiring a standing wave). The mechanism relies on second-harmonic generation (SHG) that involves the off-diagonal elements of the second-order susceptibility tensor $\hat{\chi}^{(2)}$. This allows the scattering currents to be induced along the propagation direction of the incident light, forming current configurations that cannot easily be excited using linear optical effects (an idea similar to that proposed in Ref.~\cite{jin21}). Using this mechanism, the anapole condition can be reached by changing the polarization angle of the incident field, without the need for implementing structural or spectral tuning. Due to the off-resonant character of the proposed anapoles, they can be realized in a wide range of frequencies in structures having different sizes and made of different materials. We demonstrate suppression of far-field radiation by nearly four orders of magnitude for a structure the size of half the wavelength made of gallium phosphide (GaP) in PMMA. Our approach opens a way to practical realization of perfectly isolated states of light, i.e., electromagnetic excitations that at a given frequency exist only in a compact region of space and nowhere else. This fundamentally appealing concept could find applications in optics, e.g., in optical quantum computers, enabling realization of hybrid light-matter states immune to decoherence.~\cite{zagoskin15,stenishchev24,basharin26}\\\,\\

\section{\label{sec:design}Design of ideal anapoles using exact current multipole expansion}

Let us start by providing a brief overview of the exact current multipole expansion. The generalized formula for the current multipole moments of a finite-sized scatterer embedded in a homogeneous dielectric medium reads~\cite{kolkowski26}:
\begin{equation}\label{eq:M}
M^{(l)}_{\alpha\beta\gamma\delta\cdots}=\frac{i}{\omega}\frac{(2l-1)!!}{(l-1)!}\int\limits_{-\infty}^{\infty}\frac{j_{l-1}(kr)}{(kr)^{l-1}}J_{\alpha}(\mathbf{r})r_{\beta}r_{\gamma}r_{\delta}\cdots d^3 \mathbf{r},
\end{equation}
where $l$ is the multipole order, $\alpha$, $\beta$, $\gamma$, $\delta$, $\dots$ $\in\{\hat{\mathbf{x}},\hat{\mathbf{y}},\hat{\mathbf{z}}\}$ are the directional indices of the Cartesian basis (written simply as $x$, $y$, and $z$ in the further text), $\omega$ is the frequency, $(\cdot)!!$ is the double factorial, $j_{l-1}(kr)$ are the spherical Bessel functions of the first kind of order $l-1$, $r_{\beta}$, $r_{\gamma}$, $r_{\delta}$, $\dots$ $\in\{x,y,z\}$ are the Cartesian coordinates, and $J_{\alpha}(\mathbf{r})$ is the Cartesian $\alpha$-component of the scattering current density $\mathbf{J}(\mathbf{r})$ defined as
\begin{equation}\label{eq:J}
\mathbf{J}(\mathbf{r})=-i\omega\epsilon_0[\epsilon_r(\mathbf{r})-\epsilon_s]\mathbf{E}(\mathbf{r}).
\end{equation}
Here, $\epsilon_r(\mathbf{r})$ is the relative electric permittivity dependent on the position $\mathbf{r}$, $\epsilon_s$ is the relative electric permittivity of the surrounding homogeneous medium, and $\mathbf{E}(\mathbf{r})$ is the total electric field. Equation~(\ref{eq:M}) is valid for scatterers of an arbitrary size and shape and allows evaluating the current multipole moments of an arbitrary order.  Using the linear mapping relations derived in Refs.~\cite{kolkowski26,kolkowski25}, one can calculate the classical (field-based) multipole expansion coefficients $a_{\text{E}/\text{M}}(l,m)$ from the current multipole moments obtained using Eqs.~(\ref{eq:M}) and (\ref{eq:J}); here, ``$\text{E}$'' and ``$\text{M}$'' stand for the electric and magnetic multipoles, respectively, and $l$ and $m$ are the expansion orders. Note that current multipoles are not divided into electric and magnetic, as opposed to classical multipoles. Using the classical expansion coefficients, one can calculate the scattering cross section as
\begin{equation}\label{eq:Qscat}
Q_{\text{scat}}=\frac{\pi}{k^2}\sum\limits_{l=1}^{\infty}\sum\limits_{m=-l}^{+l}(2l+1)\left[|a_{\text{E}}(l,m)|^2+|a_{\text{M}}(l,m)|^2\right].
\end{equation}
This provides a direct means to quantify the radiation of an arbitrary localized distribution of the scattering current density. Note that the above expression gives the absolute scattering cross section expressed in the units of surface area, as opposed to the frequently used normalized cross section~\cite{kolkowski26}, which is obtained by dividing $Q_{\text{scat}}$ by the geometric cross section of the scatterer.

Let us now consider the current configuration shown in Fig.~\ref{fig:1}. The only nonzero component of the current density is $J_z(\mathbf{r})$. It forms seven localized current elements $I_iL$: one at the origin of the coordinate system ($I_{\text{A}}L$) and six displaced along the $x$, $y$, and $z$-direction by distance $r_0$ ($I_{\text{B}}L$, $I_{\text{C}}L$, and $I_{\text{D}}L$). We evaluate the current multipole moments of this distribution using Eq.~(\ref{eq:M}). The only nonzero current dipole moment is
\begin{equation}\label{eq:pz}
p_z=\frac{iL}{\omega}[I_{\text{A}}+2j_0(kr_0) (I_{\text{B}}+I_{\text{C}}+ I_{\text{D}})],
\end{equation}
while the nonzero higher-order moments are 
\begin{equation}
\label{eq:Mzx}
\begin{pmatrix}M^{(l)}_{zx^{l-1}}\\M^{(l)}_{zy^{l-1}}\\M^{(l)}_{z^l}\end{pmatrix}=\frac{2iL}{\omega}\frac{(2l-1)!!}{(l-1)!}\frac{j_{l-1}(kr_0)}{k^{l-1}}\begin{pmatrix}I_{\text{B}}\\I_{\text{C}}\\I_{\text{D}}\end{pmatrix}.
\end{equation}
For example, the nonzero current octupole moments are
\begin{equation}
\label{eq:Ozx}
\begin{pmatrix}O_{zx^2}\\O_{zy^2}\\O_{z^3}\end{pmatrix}=\frac{15iL}{\omega k^2}j_2(kr_0)\begin{pmatrix}I_{\text{B}}\\I_{\text{C}}\\I_{\text{D}}\end{pmatrix},
\end{equation}
the nonzero current triacontadipole (32-pole) moments are
\begin{equation}
\label{eq:Tzx}
\begin{pmatrix}X_{zx^4}\\ X_{zy^4}\\ X_{z^5}\end{pmatrix}=\frac{315iL}{4\omega k^4}j_4(kr_0)\begin{pmatrix}I_{\text{B}}\\I_{\text{C}}\\I_{\text{D}}\end{pmatrix},
\end{equation}
and the nonzero current hectaduocontaoctupole (128-pole) moments are
\begin{equation}
\label{eq:Hzx}
\begin{pmatrix}Y_{zx^6}\\Y_{zy^6}\\Y_{z^7}\end{pmatrix}=\frac{3003iL}{8\omega k^6}j_6(kr_0)\begin{pmatrix}I_{\text{B}}\\I_{\text{C}}\\I_{\text{D}}\end{pmatrix}.
\end{equation}
Here, the index of the multipole moment raised to power $i$ means that it is repeated $i$ times in the conventional expression (e.g., $O_{zx^2}=O_{zxx}$). The moments in Eqs.~(\ref{eq:pz}), (\ref{eq:Ozx}), (\ref{eq:Tzx}), and (\ref{eq:Hzx}) are the only nonzero current multipole moments up to the order $l=8$, since the even-order moments (quadrupoles, hexadecapoles, hexacontatetrapoles) vanish due to the fact that $J_z(\mathbf{r})=J_z(-\mathbf{r})$. Furthermore, all the moments containing a product of different Cartesian coordinates in the integral of Eq.~(\ref{eq:M}) are equal to zero (e.g., $O_{zxy}=0$, $X_{zx^3y}=0$, ...), since all the current elements are positioned on the Cartesian axes, meaning that at least two out of three Cartesian coordinates are always equal to zero. Note that the expressions for the multipole moments like in Eqs.~(\ref{eq:pz})-(\ref{eq:Hzx}) remain simple even at high multipole orders, which is one of the main benefits of using the current multipole expansion. Note also that $J_{\text{A}}$ contributes only to $p_z$, since we have $j_{l-1}(0)=0$ for $l>1$.  

Next, we use the mapping relations derived in Refs.~\cite{kolkowski26,kolkowski25} to calculate the classical electric and magnetic multipole moments out of the current multipole moments in Eqs.~(\ref{eq:pz})-(\ref{eq:Hzx}):
\begin{multline}\label{eq:aE10}
\qquad a_{\text{E}}(1,0)=\frac{\sqrt{2}\,ik^3}{90\pi\epsilon E_0}\left[-15p_z\right.\\ \left.+k^2\left(O_{zx^2}+O_{zy^2}-2O_{z^3}\right)\right],\qquad
\end{multline}

\begin{multline}\label{eq:aE30}
a_{\text{E}}(3,0)=\frac{\sqrt{3}\,ik^5}{2205\pi\epsilon E_0}\left[-21\left(O_{zx^2}+O_{zy^2}-2O_{z^3}\right)\right.\\ \left.+k^2\left(3X_{zx^4}+3X_{zy^4}+8X_{z^5}\right)\right],
\end{multline}

\begin{multline}\label{eq:aE32}
a_{\text{E}}(3,\pm 2)=\frac{\sqrt{10}\,ik^5}{1470\pi\epsilon E_0}\left[7\left(O_{zx^2}-O_{zy^2}\right)\right.\\\left.-k^2\left(X_{zx^4}-X_{zy^4}\right)\right],
\end{multline}

\begin{multline}\label{eq:aE50}
a_{\text{E}}(5,0)=\frac{\sqrt{30}\,ik^7}{990990\pi\epsilon E_0}\left[-429\left(X_{zx^4}+X_{zy^4}\right)-1144 X_{z^5}\right.\\\left.+75k^2\left(Y_{zx^6}+Y_{zy^6}\right)-240 Y_{z^7}\right],
\end{multline}

\begin{multline}\label{eq:aE52}
a_{\text{E}}(5,\pm 2)=\frac{\sqrt{7}\,ik^7}{165165\pi\epsilon E_0}\left[143\left(X_{zx^4}-X_{zy^4}\right)\right.\\\left.-25k^2\left(Y_{zx^6}-Y_{zy^6}\right)\right],
\end{multline}

\begin{multline}\label{eq:aE54}
a_{\text{E}}(5,\pm 4)=\frac{\sqrt{21}\,ik^7}{330330\pi\epsilon E_0}\left[-143\left(X_{zx^4}+X_{zy^4}\right)\right.\\\left.+25k^2\left(Y_{zx^6}+Y_{zy^6}\right)\right],
\end{multline}

\begin{equation}\label{eq:aM22}
a_{\text{M}}(2,\pm 2)=\mp\frac{ik^5}{30\pi\epsilon E_0}\left(O_{zx^2}-O_{zy^2}\right),
\end{equation}

\begin{equation}\label{eq:aM42}
a_{\text{M}}(4,\pm 2)=\mp\frac{\sqrt{2}\,ik^7}{630\pi\epsilon E_0}\left(X_{zx^4}-X_{zy^4}\right),
\end{equation}

\begin{equation}\label{eq:aM44}
a_{\text{M}}(4,\pm 4)=\pm\frac{\sqrt{14}\,ik^7}{630\pi\epsilon E_0}\left(X_{zx^4}+X_{zy^4}\right),
\end{equation}

\begin{equation}\label{eq:aM62}
a_{\text{M}}(6,\pm 2)=\mp\frac{\sqrt{10}\,ik^9}{12012\pi\epsilon E_0}\left(Y_{zx^6}-Y_{zy^6}\right),
\end{equation}

\begin{equation}\label{eq:aM64}
a_{\text{M}}(6,\pm 4)=\pm\frac{\sqrt{3}\,ik^9}{3003\pi\epsilon E_0}\left(Y_{zx^6}+Y_{zy^6}\right),
\end{equation}

\begin{equation}\label{eq:aM66}
a_{\text{M}}(6,\pm 6)=\mp\frac{\sqrt{22}\,ik^9}{4004\pi\epsilon E_0}\left(Y_{zx^6}-Y_{zy^6}\right),
\end{equation}
where $\epsilon=\epsilon_0\epsilon_s$ is the electric permittivity of the surrounding medium and $E_0$ is the amplitude of the incident field that induces the scattering current density $J_z(\mathbf{r})$. Since we consider a predefined distribution of $J_z(\mathbf{r})$, it is useful to remove $E_0$ from the expressions by normalizing the total scattering cross section by the cross section of a single current element positioned at the origin (0,0,0) and having amplitude $I_{\Sigma}$ equal to 
\begin{equation}
I_{\Sigma}=|I_{\text{A}}|+2|I_{\text{B}}|+2|I_{\text{C}}|+2|I_{\text{D}}|.
\end{equation}
The normalized scattering cross section is then expressed as
\begin{equation}\label{eq:Qnorm}
\tilde{Q}_{\text{scat}}(I_{\text{A}},I_{\text{B}},I_{\text{C}},I_{\text{D}})=\frac{Q_{\text{scat}}(I_{\text{A}},I_{\text{B}},I_{\text{C}},I_{\text{D}})}{Q_{\text{scat}}(I_{\Sigma})},
\end{equation}
where $Q_{\text{scat}}(I_{\Sigma})$ can be obtained by using Eqs.~(\ref{eq:Qscat}), (\ref{eq:pz}), and (\ref{eq:aE10}) by setting the current amplitudes at positions $(\pm r_0, 0, 0)$, $(0, \pm r_0, 0)$, and $(0, 0, \pm r_0)$ to zero:
\begin{equation}\label{eq:QJSigma}
Q_{\text{scat}}(J_{\Sigma})=\frac{ k^4L^2I_{\Sigma}^2}{6\pi\epsilon^2 E_0^2 \omega^2}.
\end{equation}
The normalization performed in Eq.~(\ref{eq:Qnorm}) renders $\tilde{Q}_{\text{scat}}(I_{\text{A}},I_{\text{B}},I_{\text{C}},I_{\text{D}})$ independent of $E_0$ and $L$. In the context of anapoles, $\tilde{Q}_{\text{scat}}(I_{\text{A}},I_{\text{B}},I_{\text{C}},I_{\text{D}})$ quantifies the suppression of the radiated power in a configuration formed by $I_{\text{A}}$, $I_{\text{B}}$, $I_{\text{C}}$, and $I_{\text{D}}$ relative to the power radiated by a dipole formed by a single current element with amplitude $I_{\Sigma}$ equivalent to the total current in the configuration.

\begin{table*}[t]
\caption{\label{tab:table1} Anapole conditions and scattering suppression for the three selected scatterers of various sizes illustrated in the insets of Figs.~\ref{fig:2}a, c, and e.}
\begin{ruledtabular}
\begin{tabular}{llcccccc}
 & &\multicolumn{3}{c}{Anapole condition}&\multicolumn{3}{c}{$\tilde{Q}_{\text{scat}}$ at the exact anapole condition; scatterer size:}
\vspace{1mm}\\
 Name&Configuration&Exact\footnote{Equivalent to $15p_z=k^2 \left(O_{zx^2}+O_{zy^2}-2O_{z^3}\right)$, Eq.~(\ref{eq:eac1}).\vspace{0.5mm}}
&Conventional\footnote{Equivalent to $p_z^{\text{approx}}=-ikT_z^{\text{approx}}$, Eq.~(\ref{eq:aac2}).\vspace{0.5mm}}&Point&$\qquad$Large\footnote{Wavelength-sized: $kr_0=\pi$, $2r_0=\lambda$.\vspace{0.5mm}}$\qquad$&Medium\footnote{Fraction of wavelength: $kr_0=\pi/3$, $2r_0=\lambda/3$.\vspace{0.5mm}}&Small\footnote{Deeply subwavelength: $kr_0=\pi/10$, $2r_0=\lambda/10$.\vspace{0.5mm}}
\vspace{1mm}\\ \hline\vspace{-3mm}\\
 Flat anapole& $I_{\text{C}}=I_{\text{D}}=0$, $I_{\text{B}}\neq 0$  & $I_{\text{B}}=\frac{I_{\text{A}}}{j_2(kr_0)-2j_0(kr_0)}$ & $I_{\text{B}}=\frac{5I_{\text{A}}}{2k^2r_0^2-10}$ & $I_{\text{B}}=-\frac{I_{\text{A}}}{2}$ & 0.374 & 7.7$\times 10^{-3}$ & 5.9$\times 10^{-5}$ \vspace{1mm}\\
 Improved anapole& $I_{\text{C}}=I_{\text{B}}\neq 0$, $I_{\text{D}}=0$ & $I_{\text{B}}=\frac{I_{\text{A}}}{2j_2(kr_0)-4j_0(kr_0)}$ & $I_{\text{B}}=\frac{5I_{\text{A}}}{4k^2r_0^2-20}$ & $I_{\text{B}}=-\frac{I_{\text{A}}}{4}$ & 0.054 & 8.9$\times 10^{-4}$ & 7.0$\times 10^{-6}$ \vspace{1mm}\\
 Ideal anapole& $I_{\text{B}}=I_{\text{C}}=I_{\text{D}}\neq 0$   & $I_{\text{B}}=-\frac{I_{\text{A}}}{6j_0(kr_0)}$ & $I_{\text{B}}=\frac{I_{\text{A}}}{k^2r_0^2-6}$ & $I_{\text{B}}=-\frac{I_{\text{A}}}{6}$ & 0.022 & 2.3$\times 10^{-6}$ & 1.4$\times 10^{-10}$ \\
\end{tabular}
\end{ruledtabular}
\end{table*}

Usually, the most significant scattering contribution comes from the lowest-order classical multipole expansion coefficients. In the absence of a magnetic dipole, the dominant contribution comes from the electric dipole, determined by $a_{\text{E}}(1,0)$. To eliminate this contribution, the magnitudes of $I_{\text{A}}$, $I_{\text{B}}$, $I_{\text{C}}$, and $I_{\text{D}}$ should be chosen to satisfy the exact anapole condition for the electric-dipole scattering obtained by setting the right-hand side of Eq.~(\ref{eq:aE10}) to zero:
\begin{equation}\label{eq:eac1}
15p_z=k^2 \left(O_{zx^2}+O_{zy^2}-2O_{z^3}\right).
\end{equation}
This condition can be explicitly written in terms of $I_{\text{A}}$, $I_{\text{B}}$, $I_{\text{C}}$, and $I_{\text{D}}$ as
\begin{equation}\label{eq:eac2}
 I_{\text{B}}+I_{\text{C}}-2I_{\text{D}}\frac{j_0(kr)+j_2(kr)}{j_2(kr)-2j_0(kr) }=\frac{I_{\text{A}}}{j_2(kr)-2j_0(kr) },
\end{equation}
which is equivalent to Eq.~(\ref{eq:eac0}). In the small-scatterer approximation ($kr_0\ll1$), the combinations of spherical Bessel functions in Eq.~(\ref{eq:eac2}) can be written as
\begin{equation}\label{eq:japprox1}
j_0(kr_0)+j_2(kr_0)\approx 1-\frac{1}{10}k^2 r_0^2,
\end{equation}
\begin{equation}\label{eq:japprox2}
j_2(kr_0)-2j_0(kr_0)\approx\frac{2}{5}k^2 r_0^2-2.
\end{equation}
and the approximate anapole condition becomes
\begin{equation}\label{eq:aac1}
 I_{\text{B}}+I_{\text{C}}+\frac{k^2 r^2 - 10}{2k^2 r^2 - 10}I_{\text{D}} = \frac{5}{2k^2 r^2-10}I_{\text{A}} ,
\end{equation}
which is equivalent to the conventional anapole condition:
\begin{equation}\label{eq:aac2}
 p_z^{\text{approx}}=-ikT_z^{\text{approx}}
\end{equation}
expressed in terms of the ``primitive'' dipole moment
\begin{equation}\label{eq:pzapprox}
 p_z^{\text{approx}}=\frac{i}{\omega}\int\limits_{-\infty}^{+\infty}J_z(\mathbf{r})d^3\mathbf{r}=\frac{iL}{\omega}\left[I_{\text{A}}+2 \left(I_{\text{B}}+ I_{\text{C}}+I_{\text{D}}\right)\right]
\end{equation}
and the toroidal dipole moment
\begin{multline}\label{eq:Tzapprox}
 T_z^{\text{approx}}=\frac{k}{10\omega}\int\limits_{-\infty}^{+\infty} \left[\mathbf{r}\cdot \mathbf{J}(\mathbf{r})z-2r^2 J_z(\mathbf{r})\right]d^3\mathbf{r}\\=-\frac{Lkr_0^2}{5\omega}\left[2\left(I_{\text{B}}+I_{\text{C}}\right)+I_{\text{D}}\right].
\end{multline}
The approximate anapole condition in Eq.~(\ref{eq:aac1}) can be further simplified by setting $kr_0\rightarrow 0$, which would correspond to an ideal point scatterer. The anapole conditions at different approximation levels are summarized in Table~\ref{tab:table1} for three selected current configurations.

\begin{figure*}[t]
\includegraphics[width=170mm]{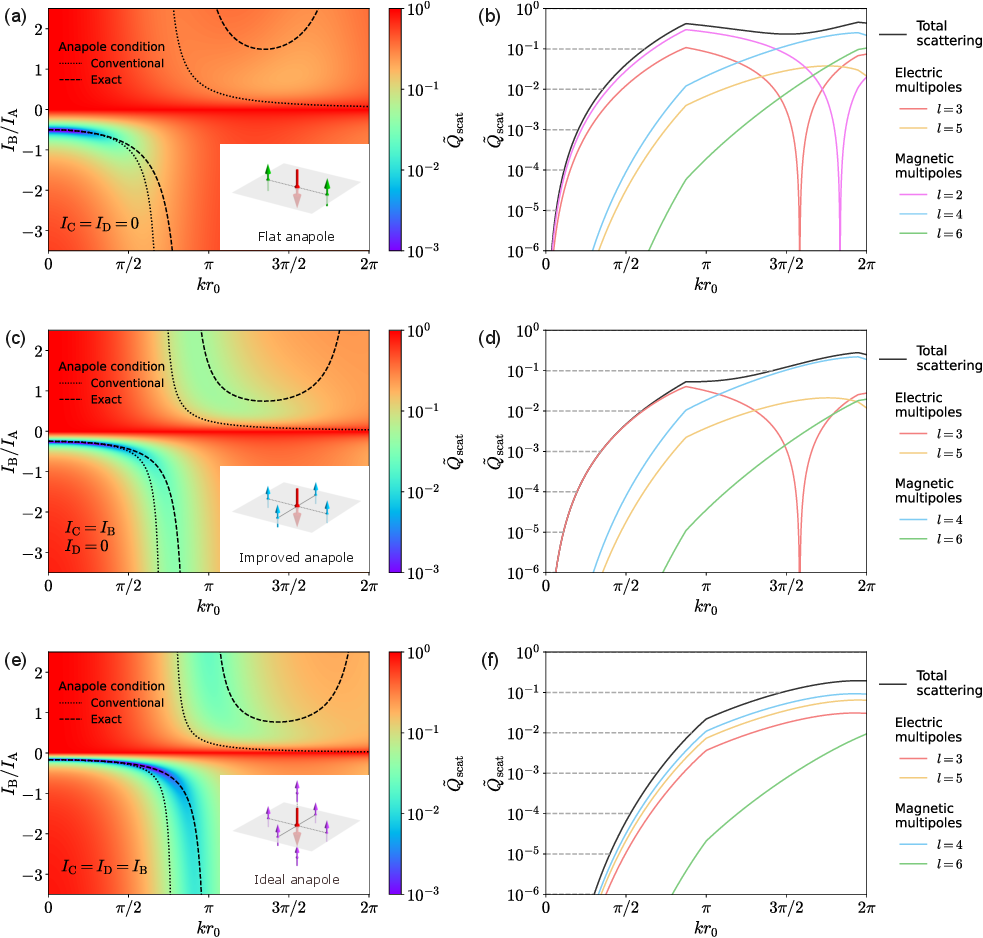}
\caption{\label{fig:2} Three selected cases of elementary current configurations and their normalized scattering cross sections $\tilde{Q}_{\text{scat}}$, which quantify the radiation suppression compared to the electric dipole radiation. The first column (a, c, e) shows $\tilde{Q}_{\text{scat}}$ as a function of $kr_0$ and $I_{\text{B}}/I_{\text{A}}$; for $kr_0=\pi$, the lateral size of the configuration $2r_0$ matches the wavelength $\lambda$ in the surrounding medium. The dashed and dotted lines indicate the parameter values satisfying the exact and conventional anapole conditions defined in Eqs.~(\ref{eq:eac1})-(\ref{eq:eac2}) and (\ref{eq:aac2}), respectively. The insets of (a, c, e) illustrate the current configurations at the anapole condition. The second column (b, d, f) shows the total $\tilde{Q}_{\text{scat}}$ (black curve) and the nonzero scattering contributions due to the classical electric and magnetic multipole moments up to the sixth order (colored curves) at the exact anapole condition. Note that $\tilde{Q}_{\text{scat}}$ is presented on a logarithmic scale in all the plots.}
\end{figure*}

Satisfying the anapole condition in Eq.~(\ref{eq:eac2}) eliminates the electric-dipole radiation, but the radiation due to the higher-order multipoles may remain significant. To minimize the total radiated power, the values of $I_{\text{B}}$, $I_{\text{C}}$, and $I_{\text{D}}$ should be further adjusted. When $I_{\text{C}}$ and $I_{\text{D}}$ are set to zero, the anapole condition leads to a configuration illustrated in the inset of Fig.~\ref{fig:2}a. This configuration approximately corresponds to a ``flat anapole'' that is commonly realized in experiments, e.g., in thin dielectric nanodiscs~\cite{miroshnichenko15} (in such cases, the excitation additionally contains octupole moments $O_{x^2z}$, forming a toroidal current together with $O_{zx^2}$). The normalized scattering cross section ($\tilde{Q}_{\text{scat}}$) for this configuration is presented in Fig.~\ref{fig:2}a as a function of $kr_0$ and the ratio $I_{\text{B}}/I_{\text{A}}$. In the small and medium-scatterer regime ($kr_0<\pi/3$ and $2r_0<\lambda/3$), both the exact and approximate anapole conditions coincide with the scattering minimum, showing scattering suppression by more than two orders of magnitude. However, for wavelength-sized scatterers ($kr_0\geq\pi$, $2r_0\geq\lambda$), the flat anapole radiates with nearly 40\% efficiency of a dipole antenna. The origin of this radiation is clearly evident from the mapping relations in Eqs.~(\ref{eq:aE30})-(\ref{eq:aM66}). Since the scattering current elements are distributed only along the $x$-axis, the current multipole moments containing the $y$ and $z$ coordinates are equal to zero ($O_{zy^2}=$ $O_{z^3}=$ $X_{zy^4}=$ $X_{z^5}=$ $Y_{zy^6}=$ $Y_{z^7}=0$). This leaves the unbalanced moments $O_{zx^2}$, $X_{zx^4}$, and $Y_{zx^6}$ present in all the coefficients in Eqs.~(\ref{eq:aE30})-(\ref{eq:aM66}). At the anapole condition, the strongest radiation is produced by the magnetic quadrupole moment $a_{\text{M}}(2,\pm 2)$, which is the lowest-order unsuppressed multipole moment for $kr_0\leq3\pi/2$ (see Fig.~\ref{fig:2}b). The values of $\tilde{Q}_{\text{scat}}$ are summarized in Table~\ref{tab:table1}.

A significant improvement in the scattering suppression is achieved by keeping $I_{\text{D}}=0$ and setting $I_{\text{C}}=I_{\text{B}}$. In this configuration, we obtain $O_{zy^2}=O_{zx^2}$, which sets $a_{\text{M}}(2,\pm 2)$ to zero and completely eliminates the magnetic-quadrupole radiation (see Eq.~(\ref{eq:aM22})). The current configuration of the improved anapole is shown in the inset of Fig.~\ref{fig:2}c, and the calculated scattering cross sections are presented in Figs.~\ref{fig:2}c and d. For wavelength-sized scatters, the radiated power at the exact anapole condition is reduced to $\sim$5\% of that produced by the equivalent electric dipole. For medium-sized and small scatterers, the level of suppression exceeds three orders of magnitude. These values constitute an improvement by one order of magnitude compared to the imperfect flat anapole considered in Figs.~\ref{fig:2}a and b. An important observation is that the conventional anapole condition clearly breaks down in this scenario, failing to correctly predict the location of the scattering minimum for $kr_0>\pi/2$. This shows that the accurate description of anapoles with improved scattering suppression requires an exact anapole condition based on the exact current multipole expansion~\cite{kolkowski26}.

After eliminating the magnetic-quadrupole radiation, the dominant contribution comes from the electric-octupole radiation due to $a_{\text{E}}(3,0)$. This contribution can be significantly reduced by setting $I_{\text{D}}=I_{\text{B}}=I_{\text{C}}$ (see Figs.~\ref{fig:2}e and f), which makes all the current octupole moments equal ($O_{zx^2}=O_{zy^2}=O_{z^3}$), leading to their mutual cancelation in $a_{\text{E}}(3,0)$, as well as in  $a_{\text{E}}(1,0)$ ($O_{zx^2}+O_{zy^2}-2O_{z^3}=0$; see Eqs.~(\ref{eq:aE10}) and (\ref{eq:aE30})). Although $a_{\text{E}}(3,0)$ is still nonzero due to the presence of current triacontadipoles ($X_{zx^4}=X_{zy^4}=X_{z^5}$), their radiation is extremely inefficient. As a result, the power radiated by a wavelength-sized configuration at the exact anapole condition is reduced to $\sim$2\% of the equivalent dipole radiation, while for a medium-sized scatterer, the suppression level approaches six orders of magnitude, as can be seen in Figs.~\ref{fig:2}e and f. The resulting \emph{ideal anapole} radiates mainly as a magnetic hexadecapole. 

Interestingly, in the range $\pi\lesssim kr_0\lesssim4\pi/3$ ($2r_0$ approximately between $\lambda$ and $4\lambda/3$), the power radiated by the improved and ideal anapoles is still equal to only a few \% of that produced by the equivalent point dipole. However, in this regime, the anapole condition requires that the peripheral current elements ($I_{\text{B}}$, $I_{\text{C}}$, and $I_{\text{D}}$) have the same sign as the element at the origin ($I_{\text{A}}$). This is in contrast to the configurations presented in the insets of Figs.~\ref{fig:2}a, c, and e, where the sign of $I_{\text{B}}$, $I_{\text{C}}$, and $I_{\text{D}}$ is opposite to that of $I_{\text{A}}$, corresponding to the anapole condition in the small-scatterer limit.

\section{\label{sec:excitation}Excitation of anapoles using second-harmonic generation}

The phases of the current elements in our designs ($I_{\text{A}}$, $I_{\text{B}}$, $I_{\text{C}}$, and $I_{\text{D}}$) either equal each other or differ by $\pi$. Excitation of such fixed-phase configurations may be challenging because, usually, the incident field is a propagating wave and its phase varies continuously along the propagation direction. The effects of phase retardation can be compensated for by engineering the scattering phase of each individual building block, following the same design principles as in phase-gradient metasurfaces~\cite{tymchenko15,li15,almeida16}. Another solution is to eliminate phase variation by using a standing wave, as proposed in Ref.~\cite{wei16} The same work also suggested that a perfectly nonradiating anapole can be excited by structuring the incident field rather than modifying the geometry of the scatterer.

Here, we propose a different approach based on nonlinear optical effects. Specifically, we consider a material with a second-order susceptibility tensor $\hat{\chi}^{(2)}$ corresponding to the zinc-blende-type crystal lattice, which is very common among technologically relevant III-V photonic materials~\cite{shoji02}. In the conventional orientation of the crystal lattice, the nonzero elements of this tensor are equal. They are the off-diagonal elements that contain three different Cartesian indices, i.e.,
\begin{equation}\label{eq:chi2xyz}
\chi^{(2)}_{xyz}=\chi^{(2)}_{xzy}=\chi^{(2)}_{yxz}=\chi^{(2)}_{yzx}=\chi^{(2)}_{zxy}=\chi^{(2)}_{zyx}=\chi^{(2)}\neq 0.\end{equation} 
In this case, the nonlinear polarization density $\mathbf{P}^{2\omega}$ at the second-harmonic (SH) frequency ($2\omega$) is equal to~\cite{sutherland19}
\begin{equation}\label{eq:P2xyz}
\mathbf{P}^{2\omega}=\begin{pmatrix}P_x^{2\omega}\\P_y^{2\omega}\\P_z^{2\omega}\end{pmatrix}=2\epsilon_0\chi^{(2)}\begin{pmatrix}E_y^{\omega} E_z^{\omega}\\E_z^{\omega} E_x^{\omega}\\E_x^{\omega} E_y^{\omega}\end{pmatrix},\end{equation} 
where $\mathbf{E}^{\omega}=(E_x^{\omega}, E_y^{\omega}, E_z^{\omega})^T$ is the field at the fundamental frequency $\omega$. If nanostructures of the nonlinear material are distributed in the $xy$-plane, they can be excited with a constant phase by a normally incident plane wave (propagating along the $z$-direction). Since such a wave may contain only the $E_x^{\omega}$ and $ E_y^{\omega}$ components, only the $P_z^{2\omega}$ component will be induced. Furthermore, if the incident wave is linearly polarized with the polarization plane making an angle $\alpha$ with the $x$-axis, i.e., $E_x^{\omega}\propto E_0^{\omega}\cos{(\alpha)}$ and $E_y^{\omega}\propto E_0^{\omega}\sin{(\alpha)}$, the amplitude and sign of $P_z^{2\omega}$ can be tuned by changing $\alpha$. For example, changing the value of $\alpha$ from positive to negative can be used to reverse the sign of $P_z^{2\omega}$. Further in this section, we show that the above properties enable straightforward realization of flat and improved anapoles confined in the $xy$-plane. Ideal anapoles can also be realized by using these properties, but due to their spatial extension in the $z$-direction, it is necessary to circumvent the effects of phase retardation by using standing-wave excitation.

\begin{figure*}[t]
\includegraphics[width=170mm]{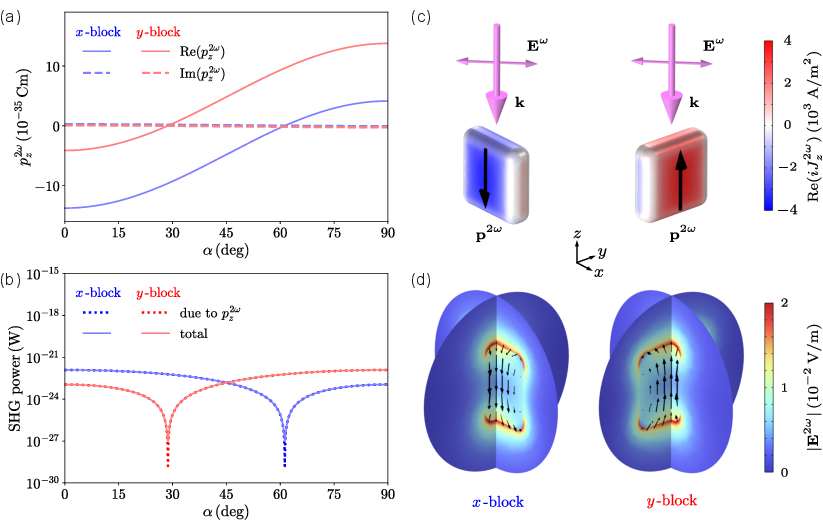}
\caption{\label{fig:3} Second-harmonic generation in anisotropic GaP blocks (20$\times$60$\times$60 nm$^3$ in size), which are used as the building blocks for nanostructures supporting excitations of the proposed anapoles. The nonlinear optical response of the blocks was modeled using COMSOL Multiphysics. (a) Real and imaginary parts (solid and dashed lines, respectively) of the current dipole moments $p_z^{2\omega}$ induced at the SH frequency in the blocks oriented along the $x$- and $y$-axes ($x$- and $y$-blocks; blue and red, respectively) by a plane wave at the fundamental frequency, $\mathbf{E}^{\omega}$, propagating along the $z$-axis and linearly polarized at an angle $\alpha$ with respect to the $x$-axis. (b) Radiated SHG power as a function of $\alpha$ for the $x$- and $y$-blocks (blue and red, respectively). The dotted lines correspond to the power radiated solely by the dipole moment $p_z^{2\omega}$, while the solid lines show the total radiated power. (c) Real part of the $z$-component of the scattering current density $\mathbf{J}^{2\omega}$ (multiplied by the imaginary unit $i$) at the surface of the $x$- and $y$-blocks (left and right, respectively) induced at the SH frequency by a plane wave linearly polarized at $\alpha$ = 45$^\circ$. The vertical pink arrows indicate the propagation direction of the incident wave, the horizontal double arrows illustrate the angle of polarization, and the black arrows on the walls of each block show the vectors of the induced current dipole moment $\mathbf{p}^{2\omega}$ (real part) at the SH frequency. (d) Spatial distributions of the electric field norm $|\mathbf{E}^{2\omega}|$ (color scale) and scattering current density $\mathbf{J}^{2\omega}$ (black arrows) under the same conditions as in (c).}
\end{figure*}

To simplify the analysis, we consider a nonlinear material with the zinc-blende crystal lattice rotated around the $z$-axis by $45^{\circ}$, such that the $[100]$ crystallographic direction lies in the $xy$-plane and is oriented at $45^{\circ}$ with respect to the $x$-axis. The modified tensor elements are~\cite{stolt26}
\begin{multline}\label{eq:chi2xyz45}
\qquad-\chi^{(2)}_{xxz}=-\chi^{(2)}_{xzx}=-\chi^{(2)}_{zxx}\\=\chi^{(2)}_{yyz}=\chi^{(2)}_{yzy}=\chi^{(2)}_{zyy}=\chi^{(2)}\neq 0,\qquad\end{multline}
and the nonlinear polarization density is
\begin{equation}\label{eq:P2xyz45}
\mathbf{P}^{2\omega}=\begin{pmatrix}P_x^{2\omega}\\P_y^{2\omega}\\P_z^{2\omega}\end{pmatrix}=\epsilon_0\chi^{(2)}\begin{pmatrix}-2E_x^{\omega} E_z^{\omega}\\2E_y^{\omega} E_z^{\omega}\\-\left(E_x^{\omega}\right)^2+\left( E_y^{\omega}\right)^2\end{pmatrix}.\end{equation} 

In nonlinear nanostructures, $\mathbf{P}^{2\omega}$ is determined by the local rather than the incident field. Hence, the nonlinear response can be engineered at nanoscale. Specifically, to enable precise control over $\mathbf{P}^{2\omega}$, we design anisotropic building blocks in which the local field amplitude is significantly enhanced for the incident field polarized along the long axis of the blocks. As a result, the relative amplitude and sign of $\mathbf{P}^{2\omega}$ depend not only on $\alpha$, but also on the orientation of the block. At the same time, the dimensions of the blocks are chosen such that the light-matter interaction remains off-resonant at both $\omega$ and $2\omega$ to prevent the local field and $\mathbf{P}^{2\omega}$ from acquiring additional phase shifts. 

We tested the above properties numerically by simulating the second-harmonic generation in realistic photonic nanostructures using the finite-element method software COMSOL Multiphysics~\cite{bachelier10,kolkowski15,carletti19,abir25}. The computational procedure is described in detail in Ref.~\cite{kolkowski26_}. It contains several steps. In the first step, the spatial distribution of the total electric field $\mathbf{E}^{\omega}$ is obtained by simulating the process of linear scattering at $\omega$, assuming that the incident field is a linearly polarized plane wave propagating along the $z$-axis. Next, $\mathbf{P}^{2\omega}$ is calculated from $\mathbf{E}^{\omega}$ using Eq.~(\ref{eq:P2xyz45}) and inserted as the source term in the Maxwell equations for a second simulation at $2\omega$, which yields the SH field $\mathbf{E}^{2\omega}$. To perform the multipole expansion at $2\omega$, the scattering current density corresponding to Eq.~(\ref{eq:J}) is calculated from
\begin{equation}\label{eq:J2}
\mathbf{J}^{2\omega}(\mathbf{r})=-2i\omega\left\{\epsilon_0[\epsilon^{2\omega}_r(\mathbf{r})-\epsilon^{2\omega}_s]\mathbf{E}^{2\omega}(\mathbf{r})+\mathbf{P}^{2\omega}(\mathbf{r})\right\}.
\end{equation}
Compared to Eq.~(\ref{eq:J}), the expression contains an additional term, $\mathbf{P}^{2\omega}(\mathbf{r})$, that accounts for the nonlinearly induced polarization density contributing to the overall field radiated at $2\omega$.

\begin{figure*}[t]
\includegraphics[width=170mm]{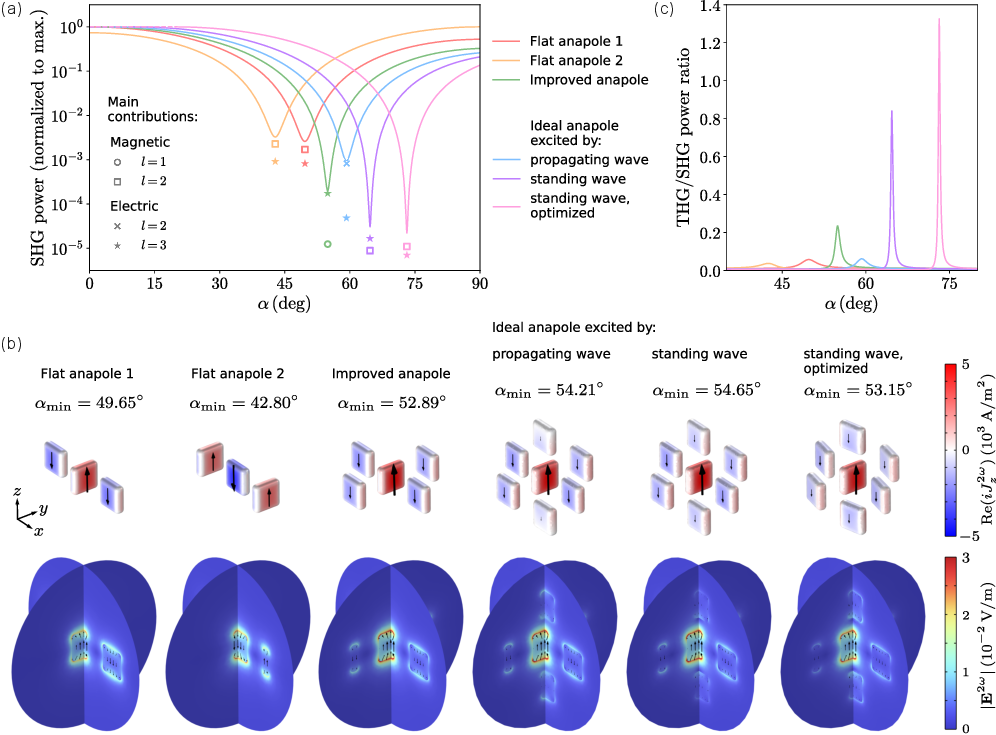}
\caption{\label{fig:4} Second-harmonic generation in photonic structures supporting flat anapoles (two cases), improved anapoles (one case), and ideal anapoles (three cases). (a) Radiated SHG power as a function of $\alpha$ for each anapole design. In each case, the values of the SHG power are normalized to the maximum value. The markers (empty square, empty circle, cross, and star) indicate the classical multipoles that have the main contributions to the radiated SHG power at an angle $\alpha=\alpha_{\text{min}}$ corresponding to the minimum of the radiated SHG power for each case (see Table~\ref{tab:table2}). (b) Spatial distributions of $iJ_z^{2\omega}$ (top row, color scale), $|\mathbf{E}^{2\omega}|$ (bottom row, color scale) and $i\mathbf{J}^{2\omega}$ (black arrows in both rows) induced at the SH frequency at $\alpha=\alpha_{\text{min}}$. (c) Ratio between the radiated THG and SHG powers as a function of $\alpha$. The peaks are located at $\alpha\approx\alpha_{\text{min}}$. For clarity, in  (a) and (c) the curves for the improved anapole (green curve) and the three cases of ideal anapoles (blue, violet, and pink) are shifted horizontally by 2$^{\circ}$, 5$^{\circ}$, 10$^{\circ}$, and 20$^{\circ}$, respectively.}
\end{figure*}

In our simulations, we consider nanostructures composed of rectangular 20-nm-wide blocks with 60 nm length and 60 nm height. Their edges and corners are rounded, with a radius of curvature equal to 8 nm. We assume that the blocks are made of gallium phosphide (GaP)~\cite{wilson20} and surrounded by PMMA. The frequency-dependent optical constants of the two materials are taken from Refs.~\cite{polyanskiy24,khmelevskaia21,beadie15}, while the $\hat{\chi}^{(2)}$ tensor of GaP is assumed to have the same form as in Eq.~(\ref{eq:chi2xyz45}). The magnitudes of $\chi^{(2)}$ and $E_0^{\omega}$ are chosen such that $\chi^{(2)}E_0^{\omega}$~=~1. The simulations are performed at a fixed fundamental frequency equal to 200 THz (vacuum wavelength of about 1500 nm), corresponding to the SH frequency of 400 THz (vacuum wavelength of about 750 nm). Since the refractive index of PMMA is approximately equal to 1.5 in the studied spectral range, the wavelength in the surrounding medium at $2\omega$ is $\lambda\approx500$~nm. Hence, the dimensions of the blocks are on the order of $\lambda/10$ and can be regarded as deeply subwavelength.

Figure~\ref{fig:3}a shows the current dipole moment $p_z^{2\omega}$ calculated using the scattering-current multipole expansion from the numerically simulated scattering current density $\mathbf{J}^{2\omega}(\mathbf{r})$ induced via SHG (see Eqs.~(\ref{eq:M}) and (\ref{eq:J2})):
\begin{equation}\label{eq:pz2}
p_z^{2\omega}=\frac{i}{2\omega}\int\limits_{-\infty}^{\infty}j_{0}(k^{2\omega}r)J_{z}^{2\omega}(\mathbf{r}) d^3 \mathbf{r}.
\end{equation}
The real and imaginary parts of $p_z^{2\omega}$ are shown as functions of the incident wave polarization angle $\alpha$ for blocks oriented along either the $x$-axis (blue) or the $y$-axis (red). At most angles, the imaginary part of $p_z^{2\omega}$ (dashed lines) is negligibly small compared to the real part (solid lines), which means that the optical response is off-resonant. At the same time, the real part of $p_z^{2\omega}$ changes from positive to negative, crossing zero at different angles $\alpha$ for the $x$- and $y$-oriented blocks. When $p_z^{2\omega}$ crosses zero, the radiated SHG power is reduced by several orders of magnitude, as shown in Fig.~\ref{fig:3}b. However, these scattering minima do not correspond to anapole excitations. They result from the fact that higher-order multipoles excited in the block in question are extremely weak when the dipole moment is equal to zero. In this case, there is essentially no radiation to be suppressed by the anapole mechanism.

The main property of interest revealed in Figs.~\ref{fig:3}a and b is the offset between the polarization angles $\alpha$ at which $p_z^{2\omega}$ crosses zero in the $x$- and $y$-blocks. This offset results from the optical anisotropy of the blocks at $\omega$ and allows us to achieve an arbitrary relative magnitude between the values of $p_z^{2\omega}$ in the two blocks. For example, Figs.~\ref{fig:3}c and d show the distributions of $\mathbf{J}^{2\omega}$ and $\mathbf{E}^{2\omega}$ in the $x$- and $y$-blocks at $\alpha=45^\circ$. In this case, $p_z^{2\omega}$ in the two blocks have the same magnitudes but opposite signs.

Understanding the nonlinear optical response of the individual blocks allows us to arrange several of them to realize the scattering current configurations corresponding to the anapoles considered in the previous section. In each case, we consider blocks that are separated by 90 nm (center-to-center), forming arrangements of lateral dimensions between 200 and 240 nm, which is about $\lambda/2$. The results obtained for different arrangements are presented in Fig.~\ref{fig:4}.

\begin{table*}[t]
\caption{\label{tab:table2} Radiated SHG power (normalized to the maximum) at $\alpha=\alpha_{\text{min}}$, main contributions of the classical multipoles to the radiated SHG power at $\alpha=\alpha_{\text{min}}$, values of $\tilde{Q}_{\text{scat}}^{2\omega}$ at $\alpha=\alpha_{\text{min}}$, and maximum THG-to-SHG ratio at $\alpha\approx\alpha_{\text{min}}$ for each of the cases presented in Fig.~\ref{fig:4}.}
\begin{ruledtabular}
\begin{tabular}{lccccccc}
 & &SHG power at $\alpha_{\text{min}}$ & \multicolumn{2}{c}{Main contributions\footnote{Contributions of the classical multipoles: MD - magnetic dipole, MQ - magnetic quadrupole, EQ - electric quadrupole, EO - electric octupole.} to SHG at $\alpha_{\text{min}}$} & $\tilde{Q}_{\text{scat}}^{2\omega}$ &THG/SHG ratio\footnote{Radiated THG power resulting from mixing the fundamental and SH fields through $\chi^{(2)}$, divided by the radiated SHG power, assuming the magnitudes of $\chi^{(2)}$ and $E_0^{\omega}$ such that $\chi^{(2)}E_0^{\omega}=1$.}
\vspace{1mm}\\
 Case&$\alpha_{\text{min}}$&(normalized to max.)& First&Second& at $\alpha_{\text{min}}$ &(max. near $\alpha_{\text{min}}$)
\vspace{1mm}\\ \hline\vspace{-3mm}\\
 Flat anapole 1& 49.65$^{\circ}$& 2.6$\times 10^{-3}$ & MQ: 65.7\% & EO: 31.6\% & 7.1$\times 10^{-3}$ & 0.058  \vspace{1mm}\\
 Flat anapole 2& 42.80$^{\circ}$& 3.3$\times 10^{-3}$ & MQ: 69.7\% & EO: 28.0\% & 7.0$\times 10^{-3}$ & 0.037 \vspace{1mm}\\
 Improved anapole& 52.89$^{\circ}$& 2.0$\times 10^{-4}$ & EO: 86.7\% & MD: 6.2\% & 8.1$\times 10^{-4}$ &  0.235 \vspace{1mm}\\
 Ideal anapole excited by:&  &  &  &   &   & &  \vspace{1mm}\\
 - propagating wave& 54.21$^{\circ}$& 9.5$\times 10^{-4}$ & EQ: 86.5\% & EO: 5.1\% & 4.2$\times 10^{-3}$ & 0.061 \vspace{1mm}\\
 - standing wave& 54.65$^{\circ}$& 3.1$\times 10^{-5}$ & EO: 54.3\% & MQ: 28.8\% & 2.2$\times 10^{-4}$ & 0.840 \vspace{1mm}\\
 - standing wave, optimized& 53.15$^{\circ}$& 2.2$\times 10^{-5}$ & MQ: 49.6\% & EO: 31.6\% &  1.7$\times 10^{-4}$ & 1.326  \\
\end{tabular}
\end{ruledtabular}
\end{table*}

For a ``flat'' anapole, there are two possible arrangements, one with the $y$-block in the center and two $x$-blocks on the sides, and the second with the $x$-block in the center and two $y$-blocks on the sides. We refer to them as ``flat anapole 1'' and ``flat anapole 2'', respectively. Figure~\ref{fig:4}a shows the radiated SHG power (normalized to its maximum value) as a function of $\alpha$ for each of the studied cases. Both types of flat anapoles (red and orange curves) enable reduction of the radiated power to 0.3\% of its maximum value. Most of the unsuppressed radiation ($\sim$65-70\%) comes from the magnetic quadrupoles, as expected for this type of anapoles (see Table~\ref{tab:table2}). The simulated distributions of $\mathbf{J}^{2\omega}$ at the SHG minima ($\alpha=\alpha_{\text{min}}$) are shown in Fig.~\ref{fig:4}b (first and second column), demonstrating that the induced currents indeed correspond to the flat anapole presented in Figs.~\ref{fig:2}a and b. The bottom row of Fig.~\ref{fig:4}b shows that the local fields at $2\omega$ remain significant despite suppression of the radiated power, which is an important property of anapoles that distinguishes them from trivial weakly radiating objects (e.g., small scatterers). 

In the case of anapoles, the capability to  confine light is usually quantified by the internal electromagnetic energy inside the scatterer. In a similar approach, one can evaluate the hypothetical maximum power that could be radiated by all the currents inside the scatterer being in phase and at a single point, like in a point dipole, and compare that to the actual radiated power. As in the previous section, we make this comparison using the relative cross section $\tilde{Q}_{\text{scat}}^{2\omega}$, which for the modeled nanostructures is calculated numerically as the ratio between the radiated SHG power and the power radiated by a point dipole of magnitude $p_{\Sigma}^{2\omega}$ calculated as
\begin{equation}\label{eq:pSigma}
p_{\Sigma}^{2\omega}=\frac{i}{2\omega}\int\limits_{-\infty}^{\infty}\left|\mathbf{J}^{2\omega}\right|d^3\mathbf{r}.
\end{equation} 
The values of $\tilde{Q}_{\text{scat}}^{2\omega}$ at $\alpha_{\text{min}}$ for both types of flat anapoles are about 7$\times 10^{-3}$ (see Table~\ref{tab:table2}).

Since we consider nonlinear optical effects, we can also calculate the power radiated at $3\omega$ due to the cascaded third-harmonic generation (THG), i.e., the sum-frequency generation resulting from mixing the fundamental and SH fields~\cite{doron19,kolkowski24}:
\begin{equation}\label{eq:P3}
\mathbf{P}^{3\omega}=\begin{pmatrix}P_x^{3\omega}\\P_y^{3\omega}\\P_z^{3\omega}\end{pmatrix}=\epsilon_0\chi^{(2)}\begin{pmatrix}-E_x^{\omega} E_z^{2\omega}-E_x^{2\omega} E_z^{\omega}\\E_y^{\omega} E_z^{2\omega}+E_y^{2\omega} E_z^{\omega}\\-E_x^{\omega}E_x^{2\omega}+ E_y^{\omega}E_y^{2\omega}\end{pmatrix}.\end{equation} 
One can expect that the THG power would also be to some extent suppressed under the anapole condition, because the scattering currents at $3\omega$ partially inherit the configuration of the SH currents. However, as the wavelength becomes shorter, the radiation due to higher-order multipoles becomes more efficient. As a result, exciting an anapole at $2\omega$ would not suppress the THG radiation as much as the SHG radiation. Therefore, the ratio between the radiated THG and SHG powers can be used as a measurable figure of merit to quantify the efficiency of light confinement at $2\omega$ due to the anapole mechanism. This THG-to-SHG ratio is plotted as a function of $\alpha$ in Fig.~\ref{fig:4}c, showing that, indeed, the excitation of anapoles gives rise to clear peaks in the plotted curves. 

Other types of anapoles introduced in Fig.~\ref{fig:2} can be analyzed in a similar way. To realize the improved anapole configuration (see Figs.~\ref{fig:2}c and d), we consider a structure that contains a single $y$-block in the center and four $x$-blocks in the peripheries (the alternative arrangement is equivalent to the structure rotated by 90$^\circ$ and therefore it does not need to be considered). The SH fields and currents at the SHG minimum for this arrangement are shown in the third column of Fig.~\ref{fig:4}b. By eliminating the magnetic-quadrupole radiation, the SHG minimum-to-maximum ratio is reduced to 0.02\% and the value of $\tilde{Q}_{\text{scat}}^{2\omega}$ will decrease to about 8$\times 10^{-4}$, which is an order-of-magnitude improvement compared to the flat anapoles (see the green curve in Fig.~\ref{fig:4}a and the third row in Table~\ref{tab:table2}). At the same time, the peak value of the THG-to-SHG ratio (see the green curve in Fig.~\ref{fig:4}c) is several times larger than the peak values for the flat anapoles. 

Apart from its remarkable performance, the unquestionable advantage of the improved anapole design is that it can be easily realized in practice by two-dimensional patterning of a thin layer of a zinc-blende-type nonlinear material and by illuminating the obtained nanostructure by a uniform linearly polarized plane wave. However, as we know from the previous section, this design can be further improved by adding two additional blocks at the top and bottom of the structure. This modification leads to a significant suppression of the electric-octupole radiation, which is responsible for 86.7\% of the SHG power radiated by the improved anapole at the SHG minimum (see Table~\ref{tab:table2}). Of course, such an ``ideal'' three-dimensional structure is much more challenging to fabricate. We find numerically that, under plane-wave excitation, the SHG minimum-to-maximum ratio and the value of $\tilde{Q}_{\text{scat}}^{2\omega}$ are 5 times larger than in the previous case (see the fourth row in Table~\ref{tab:table2}). This results from the increased size and the effect of phase retardation along the $z$-direction which gives rise to a significant electric-quadrupole scattering (contributing $86.5\%$ of the radiated SHG power at the minimum). 

To eliminate the effect of phase retardation, we choose the incident field to be a standing wave with an antinode at $z=0$. Such an incident field can be realized, e.g., by placing a mirror parallel to the $xy$-plane at a proper distance $z$ and illuminating the whole system by the same plane wave as previously. Indeed, Fig.~\ref{fig:4}a shows that the excitation of the structure by a standing wave significantly improves its SHG radiation suppression, with the SHG minimum-to-maximum ratio being reduced to 3.1$\times$10$^{-5}$ and the value of $\tilde{Q}_{\text{scat}}^{2\omega}$ to 2.2$\times$10$^{-4}$. The geometry can be further optimized by rotating the peripheral blocks to compensate for the $z$-variation of the standing wave intensity. For the optimized rotation angle of 12$^{\circ}$ counterclockwise for the top and bottom blocks and clockwise for the four lateral blocks (see the last column in Fig.~\ref{fig:4}b), the SHG minimum-to-maximum ratio is 2.2$\times$10$^{-5}$ and $\tilde{Q}_{\text{scat}}^{2\omega}$ is 1.7$\times$10$^{-4}$. Further optimization of the structure is possible by adjusting the geometry (the relative sizes and positions of the blocks). However, this would make the structure unnecessarily too demanding for practical realization. We note that, as opposed to the ideal anapoles, illuminating the flat and improved anapoles by a standing wave does not lead to a significant improvement of their radiation suppression, because the main contributions to the SHG radiation in these cases are not caused by phase retardations of the incident wave. Under the standing wave excitation, the ideal anapoles reach the highest peak values of the THG-to-SHG ratio: 3.5--5.5 times higher than for the improved anapoles and 14--38 times higher than for the flat anapoles (see the last column in Table~\ref{tab:table2}). This confirms the fact that the ideal anapoles are indeed the ultimate design for excitation of perfectly nonradiating electromagnetic fields.

\section*{Conclusions and outlook}

In this work, we have proposed and numerically demonstrated an unconventional approach to realize electromagnetic anapoles. We have shown that, by using the nonlinear optical phenomenon of second-harmonic generation, it is possible to excite configurations of localized currents that are almost perfectly dark. Specifically, we have achieved scattering suppression by nearly four orders of magnitude in structures of sizes on the order of $\lambda/2$. At the same time, by simulating the cascaded third-harmonic generation (sum frequency of the fundamental and second harmonic), we have established the THG-to-SHG ratio as a practical figure of merit for the anapole excitations at the SH frequency, which could enable their direct probing in the future experiments.

One of the main advantages of the proposed anapoles is that they can be realized by tuning the polarization angle of the incident light, which is much more convenient than changing the nanostructure geometry. Although the proposed geometry for the ideal anapoles is a three-dimensional arrangement of rectangular blocks that may be challenging to fabricate, the simpler non-ideal configurations are essentially two-dimensional and can easily be fabricated by nanopatterning a thin layer of a nonlinear material using standard lithographic methods. Importantly, the proposed anapole excitation mechanism is off-resonant and can be used in a broad spectral range, as opposed to conventional excitation of anapoles, which usually depends on the resonant response of the nanostructures and therefore occurs at a single frequency in a given structure. 

In principle, it is possible to realize the proposed anapoles also via linear excitation. This would require engineering the amplitude and phase of the individual building blocks (which is well established, e.g., for plasmonic nanoantennas) and illuminating the whole system by a standing wave formed along two Cartesian directions ($x$ and $y$). Resonant excitation would allow for the anapoles to accumulate and store optical energy (acting more like optical quasi-bound states in the continuum)~\cite{monticone19}, providing a much stronger local field enhancement than in the case of off-resonant excitation. However, simultaneous adjustment of the amplitude and phase of the locally induced dipole moments may require precise tuning of several geometric parameters at once, which may be a cumbersome task. In contrast, the nonlinear excitation proposed in this work allows the anapole condition to be reached by fine-tuning only one parameter: the polarization angle of the incident field. 

A unique property of the nonlinearly excited ideal anapoles obtained is the fact that their second-harmonic fields exist only inside the anapole structure and nowhere else. This is fundamentally different from the linear excitation, in which case the anapole fields overlap and interfere with the incident field of the same frequency. Therefore, the nonlinearly excited ideal anapoles are states of light that are truly localized in space and perfectly isolated from the environment, which makes them particularly attractive for applications in optical sensing and quantum computing~\cite{zagoskin15,stenishchev24,basharin26}.

Our work shows that electromagnetic scatterers with predefined scattering properties can be effectively designed using the recently established exact scattering-current multipole expansion~\cite{kolkowski26}. The main advantage of this theoretical framework is the intuitive character of the elementary Cartesian multipole moments. Moreover, the approach is equally applicable to small and large scattering systems of arbitrary complexities, as opposed to the conventional anapole description based on suppression of the electric dipole radiation and valid only in the small-scatterer approximation. The nonlinear excitation of ideal anapoles presented here can serve as a tutorial example, demonstrating the use of the current multipole expansion for rigorous engineering of scattering systems, both linear and nonlinear, for wide-range applications, including photonic metasurfaces and metamaterials.

\begin{acknowledgments}
R. K. has received funding from the European Union’s Horizon Europe programme for research and innovation under the Marie Sk\l{}odowska-Curie Grant Agreement No. 101060306 (NExIA). The authors acknowledge the support of the Research Council of Finland (Grants No. 347449, 353758 and 368485). For computational resources, the authors acknowledge the Aalto University School of Science “Science-IT” project and CSC – IT Center for Science, Finland.
\end{acknowledgments}

\section*{Data Availability Statement}
The data that support the findings of this study are openly available in Fairdata at \href{https://doi.org/10.23729/fd-42dfa0ac-8a07-3ec6-a33a-4ddff1985c74}{https://doi.org/10.23729/fd-42dfa0ac-8a07-3ec6-a33a-4ddff1985c74}, reference number~\cite{kolkowski26_}.

\vspace{6mm}

\nocite{*}
\bibliography{lib}
\end{document}